\documentclass{elsart}
\usepackage{graphicx}
\usepackage{amssymb}
\usepackage{amsfonts}
\usepackage{bm}

\usepackage{natbib}

\usepackage{amssymb}

\begin{document}

\begin{frontmatter}


\title{Spin-transfer physics and the model of ferromagnetism in (Ga,Mn)As}

\author{Hideo Ohno}
\address{Laboratory for Nanoelectronics and Spintronics, Research Institute of Electrical Communication, Tohoku University, Sendai 980-8577, Japan and ERATO Semiconductor Spintronics Project, Japan Science and Technology Agency, Japan, e-mail: ohno@riec.tohoku.ac.jp}

\author{Tomasz Dietl}
\address{Laboratory for Cryogenic and Spintronic Research, Institute of Physics, Polish Academy of Sciences  and  ERATO Semiconductor Spintronics Project, al. Lotnik\'ow 32/46, PL-02668 Warszawa, Poland; Institute of Theoretical Physics, Warsaw
University, PL-00681 Warszawa, Poland}





\begin{abstract}
We describe recent progress and open questions in the physics of current-induced domain-wall displacement and creep in (Ga,Mn)As. Furthermore, the reasons are recalled why, despite strong disorder and localization, the $p$-$d$ Zener model is suitable for the description of this system.
\end{abstract}


\end{frontmatter}

\label{}










\section{Introduction}
Since its first synthesis \cite{Munekata:1989_a} and the discovery of ferromagnetism \cite{Ohno:1992_a}, magnetic III-V semiconductors have been a test-bench for a number of physical concepts \cite{Ohno:1998_a,Matsukura:2002_c,Jungwirth:2006_a} ranging from revealing interacting magnetic, electrical and optical properties of the material, through testing models of ferromagnetism, to enabling device structures that uncover new spin-related phenomena. We discuss two current topics for the ferromagnetic semiconductor (Ga,Mn)As to delineate the present focus of interest and the status of understanding of this material.

\section{Spin transfer}

Owing to its small magnetization (one hundredth of that in magnetic metals), high carrier spin-polarization, and carrier-mediated ferromagnetism, the ferromagnetic III-V semiconductor (Ga,Mn)As has revealed rich spin-related physics. Such effects are either not possible to realize in metal counterparts, such as electric-field control of magnetism \cite{Ohno:2000_a}, or very difficult to access in ferromagnetic metal structures. In this section, we present a discussion of spin-transfer physics delineated by current-induced domain-wall  motion in (Ga,Mn)As and the status of its understanding.

Current-induced domain-wall motion was observed using 17 -- 20~nm thick (Ga,Mn)As strips having easy-axis perpendicular to the plane \cite{Yamanouchi:2004_a,Yamanouchi:2006_a}. The domain-wall velocity $v$ versus current-density $j$ curves were measured using a magneto-optical Kerr effect microscope and was compared with theories developed by Tatara and Kohno \cite{Tatara:2004_a} (hereafter TK), and Barnes and Maekawa \cite{Barnes:2005_a} (BM). We employed the disorder-free $p$-$d$ Zener model of ferromagnetism to calculate the domain-wall width and the spin-polarization, both of which are not yet determined experimentally but are needed to compare our experiments with the theories of domain-wall motion. This model assumes that carriers (holes; (Ga,Mn)As is a $p$-type semiconductor because Mn is an acceptor in GaAs) in disorder-free semiconductor valence bands interact with localized Mn through the exchange interaction \cite{Dietl:2000_a,Dietl:2001_b}. We discuss this model in the next section. It has been noted that:
\begin{enumerate}
\item The magnitude of the approximately linear slope of the  $v$ vs. $j$ dependence (Figure 1 (a)) is consistent within a factor of two (experimental value twice as large as the theoretical estimate) with 100\% spin transfer from the spin-polarized carriers to the localized Mn moments in the domain-wall.
\item The magnitude and temperature dependence of the threshold current density $j_{\mathrm{C}}$ is consistent with the intrinsic $j_{\mathrm{C}}$, as given by TK. In particular, the temperature dependence of $j_{\mathrm{C}}$ is closely proportional to the temperature dependence of magnetization of (Ga,Mn)As. This is consistent with TK as the intrinsic $j_{\mathrm{C}}$ is proportional to the ratio of hard-axis anisotropy (proportional to the square of the magnetization) to current spin polarization (assumed to be proportional to thermodynamic carrier spin polarization that in turn follows the magnetization in the experimentally relevant temperature range).
\item The subthreshold slope is employed to evaluate the magnitude of the so-called $\beta$-term \cite{Zhang:2004_a}, assuming that it is present in this system, and shown to be of the order of 0.01.
\end{enumerate}

\begin{figure}
\includegraphics[scale=0.6]{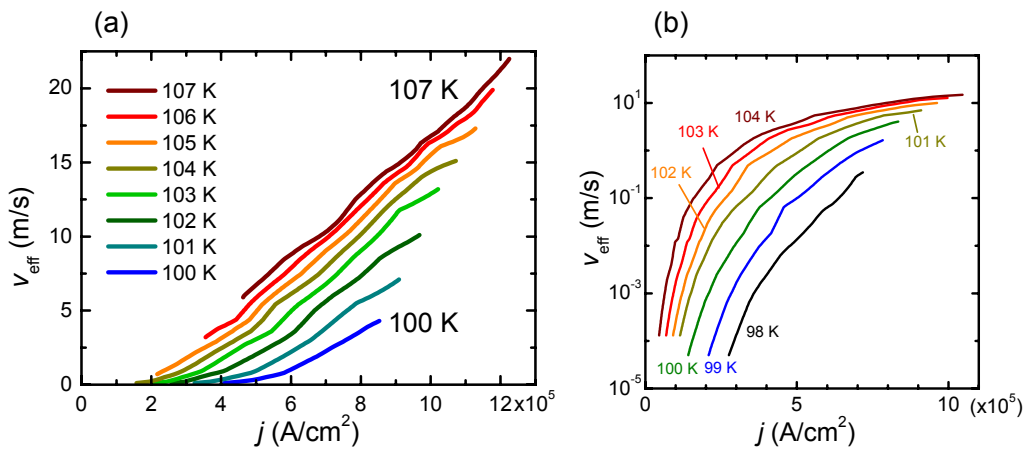}
\caption{(a) Domain-wall velocity as a function of current density. (b) Domain-wall velocity (in logarithmic
scale) as a function of current density showing slow domain-wall motion at low current density. For details see references 8 and 25.}
\end{figure}

There are several unanswered questions that need to be addressed in the future studies.

The first is  the effect of {\em the spin-orbit interaction}. This has been recently addressed theoretically in the ballistic regime \cite{Nguyen:2007_a} but not much is known in the experimentally relevant diffusive regime. In the absence of spin-splitting, the valence band states of GaAs, and thus (Ga,Mn)As, are split into three doubly degenerate states: the heavy-hole band, the light hole-band and the split-off band. The separation of the last band from the first two at the zone center is a direct measure of the spin-orbit interaction and is 0.3~eV in GaAs. When magnetization sets in, the degeneracy is further lifted so we have six spin-split bands. Currently, there is a lack of knowledge about the spin-transfer effects in the flow of spin-orbit coupled carriers. A simple picture is to think about carriers having 3/2 spin rather than 1/2. This reduces the discrepancy of the spin-transfer rates observed in Fig.~1(a), although we prefer not to go too far with this idea as there are a number of assumptions made to obtain the theoretically estimated spin-transfer rate that was compared with experiment.

The second is  {\em the intrinsic $j_{\mathrm{C}}$}. According to TK, the spin transfer torque has to overcome the conteracting torque generated by the anisotropy field, giving rise to the intrinsic $j_{\mathrm{C}}$. On the other hand, the theoretically expected temperature dependence of the extrinsic $j_{\mathrm{C}}$ due to pinning is not available. It may also be related to the ratio of the hard axis anisotropy to the current polarization \cite{Barnes}, which then makes it not possible to distinguish the two models based on the experimental knowledge we have at the moment. It is perhaps worth mentioning that we have not observed current-induced domain-wall motion in samples having an in-plane easy axis, as pointed out by Tang {\em et al.} \cite{Tang:2006_a}. This gives further support to the involvement of the hard-axis anisotropy in determining the threshold current density. This is because from separate measurements we expect the hard-axis anisotropy in samples with the easy-axis in-plane to be an order of magnitude higher than the out-of-plane anisotropy. The resulting on order of magnitude higher threshold current density is experimentally difficult to access in semiconductor samples having higher resistivities than ferromagnetic metals.

The third is the issue of {\em the field-like $\beta$ term} (for discussion of the theory aspects see \cite{Tserkovnyak} in this issue). In order to explain the current-induced domain-wall motion in metals, Zhang and Li \cite{Zhang:2004_a} invoked a field-like term that is perpendicular to the spin-transfer term. It acts on the localized spins in a domain-wall as an effective magnetic field and arises when moderate spin flip is involved in the spin-transfer process \cite{Zhang:2004_a}. This term is a subject of a number of theoretical as well as experimental studies now underway to probe its presence and quantify it  \cite{Thiaville:2005_a,Tulapurkar:2005_a,Xiao:2006_a,Tserkovnyak:2006_a,Sankey:2007_a,Seo:2007}. TK also invoked that carrier scattering by domain-walls, which gives rise to the domain-wall resistance, could result in a field-like term. Using the TK formula together with our measurement on domain-wall resistance \cite{Chiba:2006_a}, we obtain effective magnetic field values that are too high to explain our experimentally determined velocities.

To address the effect of the field-like term, measurements and analysis of domain-wall velocities in the creep regime have recently been carried out. This regime is below the threshold current density, but competition between the driving force and disorder results in a slow activated motion of the domain-wall (Figure 1 (b)). The idea is that if the field-like term is present, it should be possible to observe a creep phenomenon similar to the one induced by an external magnetic field, as reported in metal ferromagnets \cite{Lemerle:1998_a}. As expected, domain-wall creep was observed in (Ga,Mn)As in both current-driven and field-driven cases. Moreover, it was found that they obey scaling laws, having functional forms similar to the one found in the metal ferromagnet case. The scaling exponents of the two, however, are incompatible, indicating that current-driven creep is fundamentally different from field-driven creep \cite{Yamanouchi:2007_a}. Thus, in the case of (Ga,Mn)As, current-induced domain-wall motion can be understood solely by the spin-transfer effect in the entire range of velocities accessed by our experiment. So far our experiments show no evidence of the presence of the field-like term.

\section{Models of ferromagnetism and metal-insulator transition}
\subsection{Disorder-free $p$-$d$ Zener model}

The $p$-$d$ Zener model of ferromagnetism in $p$-type, tetrahedrally-coordinated diluted magnetic semiconductors designed to describe and predict quantitatively materials' properties \cite{Dietl:2000_a} has been used to calculate properties relevant in interpreting the domain-wall motion. We, therefore, discuss the background of this model here. This model has been developed in the spirit of the second-principles method (as opposed to first-principles computations) which exploits, whenever possible, experimental information on  pertinent material characteristics, such as the host band structure, the ground state of the relevant magnetic impurities, and the coupling between these two subsystems. In particular, the host band structure can be accurately parameterized  by the multi-band $kp$ method \cite{Dietl:2001_b,Hankiewicz:2004_c}, the multi-orbital tight-binding approximation \cite{Sankowski:2005_a,Timm:2005_b,Sankowski:2007_a}, or their combination \cite{Vurgaftman:2001_a}.

A standard and experimentally verified approach for the description of semiconductor alloys such as (Al,Ga)As, involves the virtual-crystal approximation (VCA). In magnetic alloys, for example (Cd,Mn)Te, VCA is supplemented by the molecular-field approximation (MFA) \cite{Dietl:1994_a}. The essence of VCA and MFA is to restore translational symmetry  by assigning to each cation site the same mean potential and mean spin polarization, obtained by averaging the potentials and polarizations of the individual atoms (here Ga and Mn with weights determined by their relative concentrations). Once translational symmetry is restored, standard methods describing band structures of semiconductors, such as the effective mass ($kp$ method) and tight-binding approaches, can be employed. The difference between the real and mean potential gives rise to alloy and spin-disorder scattering, which are typically taken into account by second-order perturbation theory.

Within the VCA and MFA, the spin splitting of carrier bands brought about by the exchange interaction between the band $s$$p$ electrons and the $d$ electrons of transition metals is proportional to magnetization of the localized spins. With this in mind, we can see how the presence of band carriers leads to ferromagnetism. According to a reasoning put forward by Zener in the 1950s, the redistribution of carriers between spin subbands, associated with a non-zero magnetization, lowers the system energy and, therefore, generates ferromagnetism. However, spontaneous magnetization will appear if temperature is low enough to make the corresponding change in entropy  irrelevant. In semiconductor alloys in question, the exchange spin splitting is especially large for the valence band, in which the wave function of the relevant carriers -- holes -- is built of anion $p$ wave functions. Accordingly, the $p$-$d$ Zener ferromagnetism occurs in $p$-type diluted magnetic semiconductor.

The disorder-free $p$-$d$ Zener model has been successful in explaining a number of experimentally observed properties in ferromagnetic III-V semiconductors, in particular (Ga,Mn)As and (In,Mn)As. The properties are the ferromagnetic transition temperature $T_{\mathrm{C}}$ \cite{Dietl:2000_a,Dietl:2001_b,Jungwirth:2006_c}, magnetization and effective Land\'e factor \cite{Sliwa:2006_a}, Gilbert constant \cite{Sinova:2004_b}, spin-polarization of the hole liquid \cite{Dietl:2001_b,Barden:2003_a}, magnetocrystalline anisotropy with its strain and temperature dependence \cite{Dietl:2000_a,Dietl:2001_b,Sawicki:2004_d,Takamura:2002_a}, magnetic stiffness \cite{Konig:2001_a,Potashnik:2002_a}, stripe domain width \cite{Dietl:2001_c}, the anomalous Hall effect \cite{Jungwirth:2006_a}, magnetic anisotropy of the Coulomb blockade \cite{Wunderlich:2006_a}, and optical properties, including the peculiar temperature dependence of the band-edge behavior of the magnetic circular dichroism \cite{Dietl:2001_b}.

\subsection{Effects of disorder and localization}

By incorporating disorder within the Boltzmann-Drude formalism it has become possible to describe the anisotropic magnetoresistance \cite{Jungwirth:2006_a} and resistance shoulder at $T<T_{\mathrm{C}}$ \cite{Lopez-Sancho:2003_b} as well as, in a semi-quantitative fashion,  the $a.c.$ conductivity \cite{Jungwirth:2006_a,Hankiewicz:2004_c}  and  domain-wall resistance \cite{Chiba:2006_a} in (Ga,Mn)As.

However, similarly to other doped semiconductors, the Boltzmann-Drude formulation is only valid in the limit where the average distance between carriers is much smaller than the relevant Bohr radius. Otherwise both quantum interference effects -- Anderson localization -- and correlation effects -- Mott localization -- are  important and can drive a metal-to-insulator transition \cite{Belitz:1994_a}. However, this Anderson-Mott transition may be viewed also as delocalization of carriers residing in an impurity band. Indeed, in the limit of small impurity concentrations, the carriers stay bound to their parent impurities. Here transport proceeds {\em via} thermally activated carrier hopping between localization centers. However, with the increase of the impurity concentration, an overlap between the carrier wave functions increases, which ultimately leads to an insulator-to-metal transition. Also within this picture, disorder and correlation effects are of primary importance, determining the impurity occupation (single {\em vs.} double) as well as the form of the density of states.

Many experiments show that (Ga,Mn)As films very often reside on or at the vicinity of the metal-insulator transition boundary \cite{Matsukura:1998_a,Jungwirth:2007_a,Sheu:2007_a}, indicating that both disorder and carrier-carrier correlations are important in this system.  A rather strong disorder originates from Coulomb potentials of acceptors, that supply holes, and donors, whose concentration is usually enhanced by a self-compensation mechanism, in addition to the local potentials introduced by magnetic impurity cores, containing spin-dependent (exchange) and spin-independent (chemical shift) contributions. The question of whether the holes accounting for ferromagnetism occupy an impurity band or reside in the GaAs-like valence band is a subject of recent discussion \cite{Jungwirth:2007_a}.

It is  a formidable task, even in non-magnetic semiconductors, to describe quantitatively effects of both disorder and carrier-carrier correlation near the Anderson-Mott transition \cite{Belitz:1994_a}. A  large number of studies in the past have led to the conclusion that doped semiconductors  near the metal-insulator transition exhibit duality, {\em i.~e.}, they show metallic band-like nature in one type of measurements, whereas at the same time they can exhibit impurity-band-like nature in another.  There were observations of 1$s$-to-2$p$ impurity transitions in metallic $n$-GaAs \cite{Liu:1993_a}, valley splitting effects in $n$-Ge \cite{Rosenbaum:1990_a}, and the presence of a Curie-Weiss component in the magnetic susceptibility \cite{Paalanen:1988_a}. All of these can be interpreted as evidence of transport taking place in the impurity band. On the other hand, weak localization theory \cite{Altshuler:1985_a,Lee:1985_a} which assumes that localization has Anderson's character, {\em i.~e.} localization of band carriers by scattering, is successful in quantitatively explaining the temperature and field dependence of the conductance. Specific heat and Pauli susceptibility measurements, for example in $n$-Si, can be explained by assuming a band mass across the metal-insulator transition \cite{Paalanen:1988_a}. In this situation, the two-fluid model of electronic states in doped semiconductors has been proposed \cite{Paalanen:1991_a}, as emphasized in papers on $p$-$d$ Zener ferromagnetism \cite{Dietl:2000_a,Dietl:2001_b}. In a somewhat simplified picture, the co-existence of two kinds of behavior results from the fact that randomness allows for the presence of isolated impurities, whose strong Curie-like paramagnetism and optical response dominate in some temperature and spectral regions, respectively, even if their concentration is statistically irrelevant.

$T_{\mathrm{C}}$ of (Ga,Mn)As, one of the thermodynamic characteristics,  shows no critical behavior at the metal-insulator transition \cite{Matsukura:1998_a}. $T_{\mathrm{C}}$ does vanish rather rapidly when moving away from the metal-insulator transition into the insulator phase \cite{Matsukura:1998_a,Sheu:2007_a}, while it grows steadily with the magnitude of the conductivity on the metal side of the metal-insulator transition \cite{Matsukura:1998_a,Potashnik:2001_a,Campion:2003_b}. Guided by these observations, the band scenario has been proposed in order to describe the ferromagnetism in magnetic III-V semiconductors on  both sides of the metal-insulator transition \cite{Dietl:2000_a,Dietl:2001_b}.  Within this model, the hole localization length, which diverges at the metal-insulator transition, remains much greater than the average distance between acceptors for the experimentally important range of hole densities.  Thus, holes can be regarded as delocalized at the length scale relevant for the coupling between magnetic ions. The spin-spin exchange interactions are effectively mediated by the itinerant carriers, so that the $p$-$d$ Zener model can be applied also to the insulator side of the metal-insulator transition. Large mesoscopic fluctuations in the local value of the density-of-states are expected near the metal-insulator transition.  Ferromagnetic order develops at $T^* > T_{\mathrm{C}}$ \cite{Mayr:2002_a} in regions where the carrier liquid supports long-range ferromagnetic correlations between randomly distributed TM spins. According to this model, the portion of the material encompassing the ferromagnetic bubbles, and thus the magnitude of the spontaneous ferromagnetic moment, grows with the net acceptor concentration, extending over the whole sample volume well within the metal phase.

A number of findings support the above scenario. For samples on the insulator side of metal-insulator transition, the field \cite{Oiwa:1997_a} and temperature \cite{Sheu:2007_a} dependence of magnetization shows that even at $T \ll T_{\mathrm{C}}$ only a fraction of spins is aligned ferromagnetically. Furthermore, the presence of randomly oriented ferromagnetic bubbles leads to a resistance maximum near $T_{\mathrm{C}}$ and to the associated negative magnetoresistance. The underlying physics is analogous to that for similar anomalies, albeit at much lower temperatures, in $n$-type DMS at the localization boundary \cite{Jaroszynski:2007_a}. Microscopically, the presence of ferromagnetic bubbles gives rise to additional spin-disorder scattering of the carriers. Such scattering together with spin-orbit scattering strongly affect the Anderson-Mott localization \cite{Belitz:1994_a,Altshuler:1985_a,Lee:1985_a}. In particular, spin-disorder scattering, once more efficient than spin-orbit scattering, destroys quantum corrections to the conductivity associated with anti-localization \cite{Timm:2005_a} and with the particle-hole triplet channel \cite{Jaroszynski:2007_a}. While both will increase the resistance upon approaching $T_{\mathrm{C}}$, the latter -- resulting from disorder-modified carrier correlation -- is usually quantitatively more significant \cite{Belitz:1994_a,Altshuler:1985_a,Lee:1985_a}.  As expected, the resistance maximum tends to disappear deeply in the metallic phase \cite{Potashnik:2001_a} (where ferromagnetic alignment is uniform and quantum localization unimportant) as well as deeply in the insulator phase \cite{Sheu:2007_a} (where ferromagnetic bubbles occupy only a small portion of the sample).

In addition to the resistance maximum at $T_{\mathrm{C}}$, a sizable negative magnetoresistance and resistance increase with lowering temperature in the metallic phase and at $T \ll T_{\mathrm{C}}$ can be explained in terms of quantum corrections to the conductivity in the weakly localized regime \cite{Matsukura:2004_a,Honolka:2007_a} for the spin-polarized universality class \cite{Dugaev:2001_a,Wojtowicz:1986_a}. Importantly, the quantum localization effects are expected to diminish gradually the low-frequency Drude spectral weight on approaching the metal-insulator transition \cite{Altshuler:1985_a,Lee:1985_a}, as indeed is observed in the {\em a.~c.} conductivity measurements \cite{Nagai:2001_a,Burch:2006_a}, and phenomenologically described as an increase in the carrier effective mass \cite{Burch:2006_a}. It is interesting to note that these measurements reveal some features suggestive of Mn impurity-related  transitions \cite{Nagai:2001_a,Burch:2006_a}, which is considered a manifestation of the two-fluid scenario \cite{Nagai:2001_a}.

\subsection{Impurity-band models}

Because of the importance of localization phenomena, a variety of models for ferromagnetism have been proposed -- but not yet  quantitatively compared with experimental findings -- which assume that in the relevant hole concentration range, the impurity and valence bands have not yet merged, so that the Fermi level actually resides within the impurity band in (Ga,Mn)As \cite{Sheu:2007_a,Inoue:2000_a,Litvinov:2001_a,Berciu:2001_a,Kaminski:2002_a}. A number of arguments summarized recently \cite{Jungwirth:2007_a} show that this is likely not the case. Without repeating those lines of reasoning, we note that the generic expectation of all the impurity models is a maximum of $T_{\mathrm{C}}$ at a half filling of the impurity band \cite{Popescu:2007_a}. This prediction is at variance with the available experimental data that show a monotonic increase of the $T_{\mathrm{C}}$ value as compensation diminishes to zero \cite{Jungwirth:2006_c,Potashnik:2001_a}. However, when the carrier density becomes greater than the magnetic impurity concentration, a decrease of the $T_{\mathrm{C}}$ value is expected as a result the RKKY-like oscillations \cite{Dietl:2001_b,Timm:2005_b}.

In view of the above arguments, we expect that low-temperature specific heat measurements will reveal the band density-of-states effective mass, which for the hole concentration, say, $10^{21}$~cm$^{-3}$ is $m^*_d \approx 1.5m_o$ in (Ga,Mn)As.

\section{Concluding remarks}

Since the domain-wall is substantially wider than the  mean free path of holes in (Ga,Mn)As \cite{Chiba:2006_a,Oszwaldowski:2006_a}, theory combining the effects of a complex band structure and disorder is needed to fully describe the domain-wall resistance and current-induced domain-wall displacement. It remains to be determined to what extent spin-transfer is affected by thermodynamic fluctuations of the magnetization as well as by inhomogeneities in the carrier distribution specific to the vicinity of the metal-to-insulator transition.

In our discussion, a random but macroscopically uniform distribution of Mn atoms has been assumed. This in not always the case in many DMS's. Spinodal decomposition of DMS films into alternating regions with low and high concentration of magnetic ions appears to account for high temperature ferromagnetism of many crystallographically uniform DMS systems \cite{Kuroda:2007_a,Katayama-Yoshida:2007_a,Dietl:2007_b}. These coherent nanocrystals may also constitute domain-wall pinning centers, as demonstrated by magnetization studies of (Ga,Mn)As films containing them \cite{Wang:2006_a}.

Another interesting question concerns the accuracy of virtual-crystal (VCA) and molecular-field approximations (MFA). A moderate magnitude of the Mn acceptor binding energy together with the experimentally estimated GaAs/(Ga,Mn)As band offset and the $p$-$d$ exchange energy indicates that VCA and MFA are valid in (Ga,Mn)As, similarly to the case of the corresponding II-VI alloys, such as (Cd,Mn)Te and (Zn,Mn)Se. However, it has recently been shown that this may not be the case in magnetically doped III-V nitrides and II-VI oxides, where a large $p$-$d$ hybridization makes the core potential of the transition metal impurities strong enough to bind a hole \cite{Dietl:2007_a}. This makes these two families of DMS similar to other non-VCA alloys, such as Ga(As,N) \cite{Wu:2002_a}. A number of surprising effects is observed in such a case, including a sign reversal of the valence-band splitting accompanied by a reduction in its magnitude \cite{Pacuski:2007_a}. The strong coupling shifts the metal-insulator transition to higher hole concentrations. However, it is not yet clear whether material characteristics will be qualitatively modified once the metallic phase is reached.

It is well known \cite{Hafner:2006_a,Rinke:2007_a} that transition metal compounds constitute a challenge for first principles methods, primarily because computational schemes derived from  density functional theory (DFT) have a tendency to predict too strong $p$-$d$ hybridization, too small on-site correlation, and too high a position of the $d$-levels, the latter revealed also by experimental studies on (Ga,Mn)As \cite{Rader:2004_a}. These difficulties are enhanced in DMS by Anderson-Mott localization effects, whose importance has been emphasized above and which generally escape from the description within DFT and related approaches. We may expect, however, a number of breakthroughs in {\em ab initio} methods relevant for DMS in the years to come.

\section*{Acknowledgments}
The work of  HO was supported in part by the IT Program of RR2002 from MEXT and the 21st Century COE program at Tohoku University, and carried out in collaboration with F. Matsukura, D. Chiba, M. Yamanouchi, J. Ieda, S. E. Barnes, S. Maekawa, and Y. Ohno. HO thanks G. Tatara for useful discussions. The work of TD was supported in part by the NANOSPIN E.~C. project (FP6-2002-IST-015728), by the Humboldt Foundation, and carried out in collaboration with M. Sawicki in Warsaw as well as with the groups of A. Bonanni in Linz, J. Cibert in Grenoble, B. Gallagher in Nottingham, S. Kuroda in Tsukuba, and L. W. Molenkamp in W\"{u}rzburg.


\begin{thebibliography}{99}
\expandafter\ifx\csname url\endcsname\relax
  \def\url#1{\texttt{#1}}\fi
\expandafter\ifx\csname urlprefix\endcsname\relax\def\urlprefix{URL }\fi

\bibitem{Munekata:1989_a}
H.~Munekata, H.~Ohno, S.~von {Moln\'{a}r}, A.~{Segm\"{u}ller}, L.~L. Chang,
  L.~Esaki,  Phys. Rev. Lett. 63 (1989)
  1849.

\bibitem{Ohno:1992_a}
H.~Ohno, H.~Munekata, T.~Penney, S.~{von Moln\'{a}r}, L.~L. Chang,
   Phys. Rev. Lett. 68 (1992) 2664.

\bibitem{Ohno:1998_a}
H.~Ohno,  Science 281 (1998) 951.

\bibitem{Matsukura:2002_c}
F.~Matsukura, H.~Ohno, T.~Dietl, Ferromagnetic semiconductors, in: K.~Buschow
  (Ed.), Handbook of Magnetic Materials, Vol.~14, Elsevier, Amsterdam, 2002,
  p.~1.

\bibitem{Jungwirth:2006_a}
T.~Jungwirth, J.~Sinova, J.~Ma{\v{s}}ek, J.~Ku{\v{c}}era, A.~H. MacDonald,
   Rev. Mod. Phys. 78 (2006)
  809.

\bibitem{Ohno:2000_a}
H.~Ohno, D.~Chiba, F.~Matsukura, T.~Omiya, E.~Abe, T.~Dietl, Y.~Ohno,
  K.~Ohtani,  Nature 408 (2000) 944.

\bibitem{Yamanouchi:2004_a}
M.~Yamanouchi, D.~Chiba, F.~Matsukura, H.~Ohno,  Nature 428 (2004) 539.

\bibitem{Yamanouchi:2006_a}
M.~Yamanouchi, D.~Chiba, F.~Matsukura, T.~Dietl, H.~Ohno,  Phys. Rev. Lett. 96 (2006) 096601.

\bibitem{Tatara:2004_a}
G.~Tatara, H.~Kohno,  Phys. Rev. Lett. 92 (2004) 086601.

\bibitem{Barnes:2005_a}
S.~E. Barnes, S.~Maekawa, Phys. Rev. Lett. 95 (2005) 107204.


\bibitem{Dietl:2000_a}
T.~Dietl, H.~Ohno, F.~Matsukura, J.~Cibert, D.~Ferrand,  Science 287 (2000)
  1019.

\bibitem{Dietl:2001_b}
T.~Dietl, H.~Ohno, F.~Matsukura,  Phys. Rev. B 63 (2001) 195205.



\bibitem{Zhang:2004_a}
S.~Zhang and Z.~Li, Phys. Rev. Lett. 93 (2004)
127204.

\bibitem{Nguyen:2007_a}
A.~K. Nguyen, H.~J. Skadsem, A.~Brataas,  Phys. Rev. Lett. 98 (2007) 146602.

\bibitem{Barnes}
S. Barnes and S. Maekawa, private communication.

\bibitem{Tang:2006_a}
H.~X. Tang, R.~K. Kawakami, D.~D. Awschalom, M.~L. Roukes, Phys.
  Rev. B 74 (2006) 041310.

\bibitem{Tserkovnyak}
Y.~Tserkovnyak, A.~Brataas, G.~E.~W. Bauer, this issue.

\bibitem{Thiaville:2005_a}
A.~Thiaville, Y.~Nakatani, J.~Miltat, Y.~Suzuki, Europhys. Lett. 69 (2005) 990.

\bibitem{Tulapurkar:2005_a}
A.~A. Tulapurkar, Y.~Suzuki, A.~Fkushima, H.~Kubota, H.~Maehara, K.~Tsunekawa,
  D.~D. Djayaprawira, N.~Watanabe, S.~Yuasa, Nature (London) 438 (2005) 339.

\bibitem{Xiao:2006_a}
J.~Xiao, A.~Zangwill, M.~D. Stiles, Phys. Rev. B 73~(2006) 054428.

\bibitem{Tserkovnyak:2006_a}
Y.~Tserkovnyak, H.~J. Skadsem, A.~Brataas, G.~E.~W. Bauer, Phys. Rev. B
  74 (2006) 144405.

\bibitem{Sankey:2007_a}
J.~C. Sankey, Y.-T. Cui, R.~A. Buhrman, D.~C. Ralph, J.~Z. Sun, J.~C.
  Slonczewski, arXiv:0705.4207v1.

\bibitem{Seo:2007}
S.~M. Seo, K.~J. Lee, W. Kim, T.~D. Lee, Appl. Phys. Lett. 90 (2007) 252508.


\bibitem{Chiba:2006_a}
D.~Chiba, M.~Yamanouchi, F.~Matsukura, T.~Dietl, H.~Ohno, Phys. Rev. Lett. 96 (2006) 096602.

\bibitem{Lemerle:1998_a}
S.~Lemerle, J.~{Fer\'ee}, C.~Chappert, V.~Mathet, T.~Giamarchi, P.~{Le
  Doussal}, Phys. Rev.
  Lett. 80 (1998) 849.

\bibitem{Yamanouchi:2007_a}
M.~Yamanouchi, J.~Ieda, F.~Matsukura, S.~E. Barnes, S.~Maekawa, H.~Ohno,
  unpublished.

\bibitem{Hankiewicz:2004_c}
E.~M. Hankiewicz, T.~Jungwirth, T.~Dietl, C.~Timm, J.~Sinova, Phys. Rev. B 70 (2004) 245211.

\bibitem{Sankowski:2005_a}
P.~Sankowski, P.~Kacman,  Phys. Rev. B 71 (2005) 201303.

\bibitem{Timm:2005_b}
C.~Timm, A.~MacDonald,  Phys. Rev. B 71 (2005) 155206.

\bibitem{Sankowski:2007_a}
P.~Sankowski, P.~Kacman, J.~A. Majewski, T.~Dietl,  Phys. Rev. B 75~(4) (2007) 045306.

\bibitem{Vurgaftman:2001_a}
I.~Vurgaftman, J.~Meyer, Phys. Rev. B 64 (2001) 245207.


\bibitem{Dietl:1994_a} T. Dietl in {\it Handbook on
Semiconductors} vol. 3B ed Moss T S (Amsterdam, Elsevier, 1994) p. 1251.


\bibitem{Jungwirth:2006_c}
T. Jungwirth and K. Y. Wang, J. Ma\v{s}ek, K. W. Edmonds, J\"{u}rgen K\"{o}nig, Jairo Sinova, M. Polini, N. A. Goncharuk
and A. H. MacDonald, M. Sawicki, A. W. Rushforth, R. P. Campion, L. X. Zhao, C. T. Foxon, B. L. Gallagher,  Phys. Rev. B 72 (2006) 165204.

\bibitem{Sliwa:2006_a}
C.~\'{S}liwa, T.~Dietl,  Phys. Rev. B 74 (2006) 245215.

\bibitem{Sinova:2004_b}
J.~Sinova, T.~Jungwirth, X.~Liu, Y.~Sasaki, J.~K. Furdyna, W.~A. Atkinson,
  A.~H. MacDonald, Phys. Rev. B 69 (2004) 085209.

\bibitem{Barden:2003_a}
J.~G. Barden, J.~S. Parker, P.~Xiong, S.~H. Chun, N.~Samarth, Phys. Rev. Lett. 91 (2003) 056602.

\bibitem{Sawicki:2004_d}
M.~Sawicki, F.~Matsukura, A.~Idziaszek, T.~Dietl, G.~Schott, C.~Ruester,
  C.~Gould, G.~Karczewski, G.~Schmidt, L.~Molenkamp,  Phys. Rev. B 70 (2004) 245325.

\bibitem{Takamura:2002_a}
K.~Takamura, F.~Matsukura, D.~Chiba, H.~Ohno, Appl. Phys. Lett. 81~(2002) 2590.

\bibitem{Konig:2001_a}
J.~{K\"{o}nig}, T.~Jungwirth, A.~MacDonald,  Phys. Rev. B 64 (2001)
  184423.


\bibitem{Potashnik:2002_a}
  S.~J.~Potashnik, K.~C.~Ku, R.~Mahendiran, S.~H.~Chun, R.~F.~Wang, N.~Samarth, P.~Schiffer, Phys. Rev. B 66 (2002) 012408.

\bibitem{Dietl:2001_c}
T.~Dietl, J.~{K\"{o}nig}, A.~H. MacDonald,  Phys. Rev. B 64 (2001) 241201.

\bibitem{Wunderlich:2006_a}
J.~Wunderlich, T.~Jungwirth, B.~Kaestner, A.~C. Irvine, A.~B. Shick, N.~Stone,
  K.-Y. Wang, U.~Rana, A.~D. Giddings, C.~T. Foxon, R.~P. Campion, D.~A.
  Williams, B.~L. Gallagher,  Phys. Rev. Lett. 97
  (2006) 077201.


\bibitem{Lopez-Sancho:2003_b}
M.~{L\'{o}pez-Sancho}, L.~Brey,  Phys. Rev. B 68
  (2005) 205322.


\bibitem{Belitz:1994_a}
D.~Belitz, T.~R. Kirkpatrick,  Rev. Mod. Phys.
  66~(1994) 261.


\bibitem{Matsukura:1998_a}
F.~Matsukura, H.~Ohno, A.~Shen, Y.~Sugawara, Phys. Rev. B 57 (1998) R2037.

\bibitem{Sheu:2007_a}
B.~L. Sheu, R.~C. Myers, J.-M. Tang, N.~Samarth, D.~D. Awschalom, P.~Schiffer,
  M.~E. Flatt\'e,
  arXiv:0708.1063v1.

\bibitem{Jungwirth:2007_a}
T.~Jungwirth, J.~Sinova, A.~H. MacDonald, B.~L. Gallagher, V.~Novak, K.~W.
  Edmonds, A.~W. Rushforth, R.~P. Campion, C.~T. Foxon, L.~Eaves, K.~Olejnik,
  J.~Masek, S.-R.~E. Yang, J.~Wunderlich, C.~Gould, L.~W. Molenkamp, T.~Dietl,
  H.~Ohno,  arXiv:0707.0665v2.

\bibitem{Liu:1993_a}
S.~Liu, K.~Karrai, F.~Dunmore, H.~D. Drew, R.~Wilson, G.~A. Thomas, Phys. Rev. B 48
  (1993) 11394.

\bibitem{Rosenbaum:1990_a}
T.~F. Rosenbaum, S.~Pepke, R.~N. Bhatt, T.~V. Ramakrishnan, 42 (1990) 11214.

\bibitem{Paalanen:1988_a}
M.~A. Paalanen, J.~E. Graebner, R.~N. Bhatt, S.~Sachdev, Phys. Rev. Lett. 61 (1988) 597.

\bibitem{Altshuler:1985_a}
B.~L. Altshuler, A.~G. Aronov, in: A.~L. Efros, M.~Pollak (Eds.),
  Electron-Electron Interactions in Disordered Systems, North Holland,
  Amsterdam, 1985, p.~1; H. Fukuyam, {\em ibid.}, p. 155.

\bibitem{Lee:1985_a}
P.~A. Lee, T.~V. Ramakrishnan, Rev. Mod. Phys. 57 (1985) 287.

\bibitem{Paalanen:1991_a}
M.~A. Paalanen, R.~N. Bhatt, S.~Sachdev, Physica B 169 (1991) 223.

\bibitem{Potashnik:2001_a}
S.~J. Potashnik, K.~C. Ku, S.~H. Chun, J.~J. Berry, N.~Samarth, P.~Schiffer,
   Appl. Phys. Lett. 79 (2001) 1495.

\bibitem{Campion:2003_b}
R.~P. Campion, K.~W. Edmonds, L.~X. Zhao, K.~Y. Wang, C.~T. Foxon, B.~L.
  Gallagher, C.~R. Staddon, J. Cryst. Growth 251 (2003) 311.

\bibitem{Mayr:2002_a}
M.~Mayr, G.~Alvarez, E.~Dagotto,
  Phys. Rev. B 65 (2002) 241202.

\bibitem{Oiwa:1997_a}
A.~Oiwa, S.~Katsumoto, A.~Endo, M.~Hirasawa, Y.~Iye, H.~Ohno, F.~Matsukura,
  A.~Shen, Y.~Sugawara, Solid State Commun. 103 (1997) 209.

\bibitem{Jaroszynski:2007_a}
J.~Jaroszy\'{n}ski, T.~Andrearczyk, G.~Karczewski, J.~Wr\'{o}bel, T.~Wojtowicz,
  D.~Popovi\'{c}, T.~Dietl,  Phys. Rev. B 76
  (2007) 045322.

\bibitem{Timm:2005_a}
C.~Timm, M.~E. Raikh, F.~{von Oppen}, Phys. Rev. Lett. 94 (2005) 036602.

\bibitem{Matsukura:2004_a}
F.~Matsukura, M.~Sawicki, T.~Dietl, D.~Chiba, H.~Ohno, Physica E 21 (2004) 1032.

\bibitem{Honolka:2007_a}
J.~Honolka, S.~Masmanidis, H.~X. Tang, D.~D. Awschalom, M.~L. Roukes,
   Phys. Rev. B 75 (2007) 245310.

\bibitem{Dugaev:2001_a} V. K. Dugaev, P. Bruno, and J. Barna\'s, Phys. Rev. B 64 (2001) 144423.


\bibitem{Wojtowicz:1986_a}
T.~Wojtowicz, T.~Dietl, M.~Sawicki, W.~Plesiewicz, J.~Jaroszy\'nski, Phys. Rev. Lett.
  56 (1986) 2419.

\bibitem{Nagai:2001_a}
Y.~Nagai, T.~Junimoto, K.~Ngasaka, H.~Nojiri, M.~Motokawa, F.~Matsujura,
  T.~Dietl, H.~Ohno,  Jpn. J. Appl. Phys. 40 (2001) 6231.

\bibitem{Burch:2006_a}
K.~S. Burch, D.~B. Shrekenhamer, E.~J. Singley, J.~Stephens, B.~L. Sheu, R.~K.~Kawakami, P.~Schiffer, N.~Samarth, D.~D. Awschalom, D.~N. Basov, Phys. Rev. Lett. 97 (2006) 087208.

\bibitem{Inoue:2000_a}
J.~Inoue, S.~Nonoyama, H.~Itoh,  Phys. Rev. Lett. 85 (2000) 4610.

\bibitem{Litvinov:2001_a}
V.~I. Litvinov, V.~K. Dugaev, Phys. Rev. Lett. 86 (2001) 5593.

\bibitem{Berciu:2001_a}
M.~Berciu, R.~N. Bhatt,  Phys. Rev. Lett. 87 (2001) 107203.

\bibitem{Kaminski:2002_a}
A.~Kaminski, S.~{Das {Sarma}}, Phys. Rev. Lett. 88 (2002) 247202.

\bibitem{Popescu:2007_a}
F.~Popescu, C.~{\c{S}}en, E.~Dagotto, A.~Moreo, arXiv:0705.0309v1.

\bibitem{Oszwaldowski:2006_a} R. Oszwa³dowski, J. A. Majewski, T. Dietl, Phys. Rev. B 74 (2006) 153310.

\bibitem{Kuroda:2007_a} S. Kuroda, N. Nishizawa, K. Takita, M. Mitome, Y. Bando, K. Osuch,T. Dietl, Nature Mat. 6 (2007) 440.

\bibitem{Katayama-Yoshida:2007_a} H.~Katayama-Yoshida, K.~Sato, T.~Fukushima, M.~Toyoda, H.~Kizaki, V.~A.Dinh, P.~H.~Dederichs, phys.stat. sol. (a) 204 (2007) 15.

\bibitem{Dietl:2007_b} T. Dietl, T. Andrearczyk, A. Lipi\'nska, M. Kiecana, Maureen Tay,  Yihong Wu, arXiv:0708.2476v1, Phys. Rev. B, in press.

\bibitem{Wang:2006_a}   K.-Y. Wang, M. Sawicki, K.W. Edmonds, R.P. Campion, A.W. Rushforth, A.A. Freeman, C.T. Foxon, B.L. Gallagher, T. Dietl,  Appl. Phys. Lett. 88 (2006) 022510.


\bibitem{Dietl:2007_a}
T. Dietl, cond-mat/cond-mat/0703278.

\bibitem{Wu:2002_a} J. Wu, W. Shan, and W Walukiewicz, Semicond. Sci. Technol. 17 (2002) 860.

\bibitem{Pacuski:2007_a} W. Pacuski, P. Kossacki, D. Ferrand, A. Golnik, J. Cibert, M. Wegscheider, A. Navarro-Quezada, A. Bonanni, M. Kiecana, M. Sawicki, T. Dietl, arXiv:0708.3296.

\bibitem{Hafner:2006_a} J. Hafner, Ch. Wolverton, G. Ceder, MRS Bulletin 31 (2006) 659.

\bibitem{Rinke:2007_a} P. Rinke, A. Qteish, J. Neugebauer, M. Scheffler, $\Psi k$ Newsletters, No 79 (2007) 163; available at
www.fhi-berlin.mpg.de/th/.


\bibitem{Rader:2004_a} O. Rader, C. Pampuch, A. M. Shikin, J. Okabayashi, T. Mizokawa, A. Fujimori, T. Hayashi, M. Tanaka, A. Tanaka, Kimura, Phys. Rev. B 69 (2004) 075202.



\end{thebibliography}
\end{document}